\documentclass[intlimits,twoside,a4paper]{article}

\usepackage{amsmath,amssymb}
\usepackage{graphicx}
\usepackage{wrapfig}
\usepackage{color}

\usepackage[T2A]{fontenc}
\usepackage[cp1251]{inputenc}

\usepackage[eqsecnum]{cmpj2}

\issue{2014}{17}{2}{23606}
\doinumber{10.5488/CMP.17.23606}

\title[Quantum transport equations for Bose systems]{Quantum transport equations for Bose systems taking into account
 nonlinear hydrodynamic processes}

\author[P.A.~Hlushak, M.V.~Tokarchuk]{P.A.~Hlushak, M.V.~Tokarchuk}

\address{Institute for condensed Matter Physics of the National Academy of Sciences of Ukraine \\
1 Svientsitskii St., 79011 Lviv, Ukraine}

\date{Received March 20, 2013, in final form May 14, 2014}
\authorcopyright{P.A.~Hlushak, M.V.~Tokarchuk, 2014}

\begin{document}

\maketitle

\begin{abstract}

Using the method of nonequilibrium statistical operator by Zubarev, an approach is proposed for the
description of kinetics which takes into account the nonlinear hydrodynamic
fluctuations for a quantum Bose system. Non-equilibrium statistical operator is presented which consistently describes both the kinetic and
nonlinear hydrodynamic processes.
Both a kinetic equation for the nonequilibrium
one-particle distribution function and a generalized Fokker-Planck equation for nonequilibrium
distribution function of hydrodynamic variables (densities of momentum, energy and particle number) are obtained. A structure function of hydrodynamic fluctuations in cumulant representation is
calculated, which makes it possible to analyse the generalized Fokker-Planck equation in
Gaussian and higher approximations of the dynamic correlations of hydrodynamic variables which is
important in describing the quantum turbulent processes.

 \pacs 67.40.-w, 47.37.+q

\end{abstract}

\section{Introduction}

The development of the nonequilibrium statistical theory which  takes into account
one-particle and collective physical processes
is a difficult problem in modern physics.
The separation of contributions from the kinetic and hydrodynamic fluctuations into
time correlation functions, excitation spectrum, and transport coefficients allows one to obtain
more information on physical processes at different time and spatial intervals that define
the dynamic properties of a system. A considerable success was achieved in the papers
\cite{ZubarMoroz1984,ZubMorOmelTok1991,ZubMorOmelTok1993,TokOmelKobr1998} in which the approach of
a consistent description of kinetics and hydrodynamics of classical dense gases and fluids is
proposed based on the Zubarev method of nonequilibrium statistical operator
\cite{Zubar1974,ZubarMoroz1989,ZubMorRop1996.1,ZubMorRop1996.2}. It is appropriate to apply this
approach for the study of the dynamics of quantum liquids such as
liquid helium in the normal and superfluid states.
The quantum system of Bose particles serves as a physical model in theoretical
descriptions of both equilibrium and nonequilibrium properties of liquid helium and trapped
weakly-interacting Bose gases \cite{PethcSmith2008,GrifNikZar2009}. Many books
\cite{Khalat1961,Patter1974,Glyde1994,Griff1993,KovalPelet2006} and articles are devoted to
a theoretical description of this system.

In the studies by  Morozov \cite{Moroz1983,Moroz1986} and
Lebedev, Sukhorukov and Khalatnikov \cite{LebSukhKhal1981} the theoretical approaches were proposed
to the description of nonlinear hydrodynamic fluctuations connected with the problem of calculating
the dispersion for kinetic transport coefficients and the spectrum of collective modes in the
low-frequency area of the superfluid Bose liquid. The microscopical derivation of hydrodynamic
equations of a superfluid liquid taking  dissipative processes into account
was presented by Kovalevsky, Lavrinenko, Peletminsky and Sokolovsky \cite{KovalPeletSlus1982},
where kinetic coefficients are expressed in terms of time correlation functions of
the corresponding flux operators. The generalized Fokker-Planck equation for
nonequilibrium distribution function of slow variables for quantum systems was obtained by Morozov
\cite{Moroz1981}.

The problems of building a kinetic equation for Bose systems based on the microscopic approach
were considered by Akhiezer and Peletminsky \cite{AkhezPelet1981}
and by Kirkpatrick and Dorfman \cite{KirkpDorf1985.1,KirkpDorf1985.2,KirkpDorf1985.3}.
The results of \cite{KirkpDorf1985.1,KirkpDorf1985.2}
were extended and used to describe the trapped weakly-interacting Bose gases at
finite temperatures \cite{ZarGrifNik1998,ZarNikGrif1999}.
Lauck, Vasconcellos and Luzzi \cite{LauckVasconLuzzi1990} developed
a nonlinear quantum transport theory for a far from equilibrium many-body system, which
is based on nonequilibrium statistical operator method.
The hierarchy of generalized evolution equations of dissipative processes in a Bose
fluid was derived by Madureira, Vasconcellos and Luzzi \cite{MadurVasconLuzzi1998.1,MadurVasconLuzzi1998.2}.
The proposed approach may
be suitable for a description of the transport coefficients in molecular hydrodynamics, where the
coefficients are frequency and wavelength dependent. Molecular hydrodynamics of nondegenerate
Bose gas \cite{Tserkov1990,Tserkov1990_2} and degenerate one \cite{Tserkov1992.1995,Tserkov1992.1995_2} was derived using
the method of two-temporal Green functions \cite{Zubarev1960,Tserkov1986}.

The nonequilibrium statistical operator of many-particle Bose system which consistently
describes the kinetics and hydrodynamics, was derived in \cite{VakHlTok1997,HlTok2005}.
The quantum nonequilibrium one-particle distribution function and the average value of
density of interaction potential energy have been selected as parameters of a consistent
description of the nonequilibrium state. Generalized transport equations were obtained for strongly
and weakly nonequilibrium Bose systems with separate contributions from both the
kinetic and potential energies of particle interaction.

The aim of the present paper is to construct transport equations for a quantum system
that take into account nonlinear hydrodynamic processes.
Large-scale fluctuations in a system, which are related to the nonlinear hydrodynamical processes,
play an essential role in the transition from normal to
superfluid state \cite{Onuki2004}, in the transition from laminar to turbulent flow,
and in the acoustic turbulence in superfluid helium \cite{QunTurb2009,Nemirov2006}.
Similar problems arise while describing low-frequency anomalies
in kinetic equations related to ``long tails'' of correlation functions
\cite{ResibLen1977,Kawasaki1976,ZubarMoroz1983,PeletSlus1992}.
To achieve this aim, a nonequilibrium
statistical operator, which consistently describes both the kinetic and nonlinear hydrodynamical
fluctuations in a quantum liquid, is derived. Then, a coupled set of kinetic equations are obtained for quantum
one-particle distribution function and generalized Fokker-Plank equations for the functional of
hydrodynamical variables: densities of particle number, momentum and energy.
A structure function of hydrodynamic fluctuations is calculated using a cumulant representation.
It provides a possibility to analyse the generalized Fokker-Planck equation in Gaussian and higher
approximations of dynamic correlations of hydrodynamic variables which is important in describing the
phase transitions and quantum turbulent processes.

\section{Kinetic equation for nonequilibrium Wigner function
and Fokker-  \\ Planck equation for distribution function of hydrodynamic variables}

The observable average values of energy density $\langle\hat{\varepsilon}_{\bf{q}}\rangle^t$,
momentum density $\langle\hat{P}_{\bf{q}}\rangle^t$, and particle numbers density
$\langle\hat{n}_{\bf{q}}\rangle^t$ are the reduced description parameters
of the hydrodynamical nonequilibrium state of a normal Bose liquid characterized by the
energy, momentum and mass flow processes. Operators for these physical quantities are defined
through the Klimontovich operator of the phase density of particle number $\hat{n}_{\bf q}({\bf
p})=\hat {a}^{+}_{\bf{p-\frac{q}{2}}}\hat{a}_{\bf{p+\frac{q}{2}}}$
\begin{equation}                \label{eq2.1}
 \hat {n}_{\bf q} = \frac{1}{\sqrt N }\sum\limits_{\bf{p}}
 {\hat {n}_{\bf q} ({\bf p})}, \qquad
 \hat {\bf P}_{\bf q} =
 \frac{1}{\sqrt N }\sum\limits_{\bf p} {{\bf p}\,\hat{n}_{\bf q} ({\bf p})},
 \qquad
 \hat{\varepsilon}_{\bf q}=\hat{\varepsilon }_{\bf q}^\textrm{kin}+\hat{\varepsilon }_{\bf q}^\textrm{int}\,,
\end{equation}
 where
$\hat{\varepsilon}_{\bf q}^\textrm{kin}$ and $\hat{\varepsilon}_{\bf{q}}^\textrm{int}$ are Fourier-components of
the operators of kinetic and potential energy densities
\begin{equation}            \label{eq2.2}
 \hat{\varepsilon }_{\bf q}^\textrm{kin}  = \frac{1}{\sqrt N }\sum\limits_{\bf p} {\left(
{\frac{p^2}{2m} - \frac{q^2}{8m}} \right)\hat {n}_{\bf q} ({\bf p})}, \qquad
\hat {\varepsilon }_{\bf q}^\textrm{int} = \frac{1}{2V\sqrt{N}}\sum\limits_{{\bf k,\,p,\,p'}}
 \nu (k) \hat{a}^{+}_{{\bf p} +
\frac{{\bf k} - {\bf q}}{2}} \, \hat {n}_{\bf k} ({\bf p}')
\hat {a}_{{\bf p} - \frac{{\bf k} - {\bf q}}{2}}\,.
\end{equation}

The average value of the phase density operator of particle number is equal to the nonequilibrium
one-particle distribution function $f_1({\bf{q,p}},t)=\langle\hat{n}_{\bf{q}}({\bf p})\rangle^t$,
which satisfies the kinetic equation for a quantum Bose system.

The agreement between the kinetics and hydrodynamics for dilute Bose gas does not cause any problems
because in this case the density is a small parameter. Therefore, only the quantum one-particle
distribution function $f_1({\bf{q}}, {\bf{p}};t)$ can be chosen as the parameter of a reduced
description. At a transition to quantum Bose liquids, the contribution of collective correlations,
which are described by average potential energy of interaction,  is more important than
one-particle correlations connected with $f_1 ({\bf q}, {\bf p}; t) $. Hence, in order to
consistently describe the kinetics and hydrodynamics of Bose liquid, the one-particle
nonequilibrium distribution function along with the average potential energy of interaction
should be chosen as the parameters of a reduced description \cite{VakHlTok1997,HlTok2005}.
The nonequilibrium state of such a quantum system is completely
described by a nonequilibrium statistical operator $\hat{\varrho}(t)$
which satisfies the quantum Liouville equation
\begin{equation}             \label{eq2.3}
 \frac{\partial}{\partial t}\hat{\varrho}(t)+ iL_{N} \hat{\varrho}(t) =  -
\varepsilon \left[ \hat {\varrho }(t) - \hat {\varrho }_\textrm{q}
(t)\right].
\end{equation}
The infinitesimal source $\varepsilon$ in the right-hand side of this equation breaks the symmetry of
the Liouville equation with respect to $t\to\,-t$ and selects retarded solutions
($\varepsilon\to\,+0$ after limiting thermodynamic transition). The quasi-equilibrium  statistical
operator $\hat{\varrho}_\textrm{q}(t)$ is determined from the condition of the informational entropy
extremum at the conservation of normalization condition $\mbox{Sp}\,\hat{\varrho}_\textrm{q}(t)=1$ for fixed
values of $\langle \hat{n}_{\bf{q}} ({\bf{p}})\rangle^t$ and $\langle\hat{\varepsilon}^\textrm{int}_{\bf{q}}
\rangle^t$ \cite{VakHlTok1997,HlTok2005}:
\begin{equation}            \label{eq2.4}
\hat{\varrho }_\textrm{q} (t) = \exp
\left\{
  -\Phi(t)- \sum\limits_{\bf q}
  \beta _{-{\bf q}}(t) \hat{\varepsilon}^\textrm{int}_{\bf q}  -
  \sum\limits_{\bf q,\,p} \gamma _{- {\bf q}} ({\bf p};t)
  \hat {n}_{\bf q} ({\bf p})
\right\},
\end{equation}
\noindent
where the Lagrangian multipliers $\beta_{-{\bf q}}(t)$, $\gamma_{-{\bf{q}}}({\bf{p}};t)$
are determined from the self-consistent conditions:
$$
\langle \hat{n}_{\bf q}({\bf p})\rangle^t=\langle \hat{n}_{\bf{q}}({\bf p})\rangle^t_\textrm{q}\,,
\qquad
\langle \hat {\varepsilon }^\textrm{int}_{\bf q}\rangle^t=
\langle\hat{\varepsilon}^\textrm{int}_{\bf q}\rangle^t_\textrm{q}\,.
$$
Here,
$\langle(\ldots)\rangle^t=\mbox{Sp}(\ldots)\hat{\varrho}(t)$ and
$\langle(\ldots)\rangle^t_\textrm{q}=\mbox{Sp}(\ldots)\hat{\varrho}_\textrm{q}(t)$.
The Massieu-Plank functional
\begin{equation}               \label{eq2.5}
\Phi (t) = \ln \,\mbox{Sp}\,\exp \left\{- \sum\limits_{\bf q}
  \beta _{-{\bf q}}(t) \hat{\varepsilon}^\textrm{int}_{\bf q}  -
  \sum\limits_{\bf q,\,p}  \gamma _{- {\bf q}} ({\bf p};t)
  \hat {n}_{\bf q} ({\bf p})
\right\}
\end{equation}
is determined from the normalization condition $\mbox{Sp}\,\hat{\varrho}_\textrm{q}(t)=1$.

The system of equations for the one-particle distribution function and the average density of
potential energy are strongly nonlinear\cite{VakHlTok1997,HlTok2005}. The system can be used
for a consistent description of the kinetics and hydrodynamics of both strongly and weakly
nonequilibrium states of the Bose systems. Projecting the transport equations on
the components of the vector ${\bf\Psi}({\bf{p}})=\left\{ {1,\,{\bf p},\,{p^2}/({2m})
- {q^2}/({8m})}\right\}$ yields the equations of nonlinear hydrodynamics, in which the
transport processes of kinetic and potential parts of the energy are described by two interdependent
equations. Obviously, such equations of nonlinear hydrodynamics provide more opportunities
to describe in detail the mutual transformation of kinetic and potential energies
during nonequilibrium processes in the system.

In this paper, as previously \cite{VakHlTok1997,HlTok2005}, the nonequilibrium quantum distribution
function $f_1({\bf{q}},{\bf{p}};t)=\langle \hat{n}_{\bf q}({\bf p})\rangle^t$ is chosen as a
parameter to describe one-particle correlations. However, to describe the collective processes in a
quantum system, we introduce the operator function
\begin{equation}
\label{eq2.6}
\hat {f}({\bf a}) = \int \rd{\bf x}\, {\re}^{\ri{\bf x}(\hat{{\bf a}}-{\bf a})},
\qquad \rd{\bf x}=\prod^{5}_{m=1}\prod\limits_{\bf k}\,\frac{\rd x_{m{\bf k}}}{2\pi}\,,
\end{equation}
%
where $\hat{{\bf a}}=\{\hat{a}_{1\bf{k}},\hat{a}_{2\bf{k}},\hat{a}_{3\bf{k}} \}$, \,
 $\hat{a}_{1{\bf k}} = \hat {n}_{\bf k} $,
 $\hat {a}_{2{\bf k}}= \hat {\bf P}_{\bf k}$,
 $\hat {a}_{3{\bf k}} = \hat {\varepsilon}_{\bf k}=\hat {\varepsilon }_{\bf k}^\textrm{kin}+
 \hat {\varepsilon}_{\bf k}^\textrm{int}$
are the Fourier-components of the operators of particle number, momentum and energy densities
(\ref{eq2.1}). The scalar values $a_{m{\bf{k}}}=\left\{n_{\bf{k}},\,\,{\bf{P}}_{\bf{k}},\,\,
\varepsilon_{\bf{k}}\right\}$ are the corresponding collective variables. The average values of
the operator function (\ref{eq2.6}) represent a microscopic distribution function  of hydrodynamic variables
obtained in accordance with Weyl correspondence rule from the classical distribution function
\cite{Moroz1981}
\begin{equation}           \label{eq2.7}
  f({\bf a}) = \delta({\bf A}-{\bf a})=\prod\limits_{m=1}^N
\prod\limits_{\bf{k}} \delta({A}_{m\bf{k}}-a_{m\bf{k}}),
\end{equation}
where ${\bf A}=\{A_{1\bf k}\ldots,A_{N\bf k}\}$ are the classical dynamical variables.

The average values $f_1({\bf q},{\bf p};t)=\langle \hat{n}_{\bf k}({\bf p})\rangle^t$,  $f({\bf
a};t)=\langle \hat{f}({\bf a})\rangle^t$ are calculated using the nonequilibrium statistical
operator $\hat{\varrho}(t)$, which satisfies the Liouville equation. In line with the idea of
a reduced description of the nonequilibrium state, the statistical operator $\hat{\varrho}(t)$ should
functionally depend on the quantum one-particle distribution function and on the distribution
functions of hydrodynamic variables
\begin{equation}            \nonumber
\hat {\varrho }(t) =
\hat {\varrho }\left[ f_1 ({\bf q},{\bf p};t),\,f({\bf a};t) \right].
\end{equation}
Thus, the task is to find a solution of the Liouville equation for $\hat{\varrho}(t)$ which has
the above form. To this end, we use the method of Zubarev nonequilibrium statistical operator
\cite{Zubar1974,ZubarMoroz1989,ZubMorRop1996.1,ZubMorRop1996.2}. We consider the Liouville equation
(\ref{eq2.3}) with infinitely small source. The source correctly selects retarded solutions in
accordance with the reduced description of nonequilibrium state of a system. The quasi-equilibrium
statistical operator $\hat{\varrho}_\textrm{q}(t)$ is determined in a usual way, from the condition of the
maximum informational entropy functional:
\[
S[\hat{\varrho}']=-\mbox{Sp}\left\{\hat{\varrho}'\ln\hat{\varrho}'\right\}-\sum_{\bf p}
\gamma _{- {\bf q}}({\bf p};t)
  \mbox{Sp}\left\{\hat{\varrho}'\hat {n}_{\bf q}({\bf p})\right\}-\int \rd a\,F({\bf a};t)
  \mbox{Sp}\left\{\hat{\varrho}'\hat{f}({\bf a})\right\}.
\]
Then, the quasi-equilibrium statistical operator can be written as
\begin{equation}             \label{eq2.8}
\hat{\varrho}_\textrm{q} (t) = \exp
\left\{
  -\Phi(t)- \sum\limits_{\bf q,\,p}  \gamma _{- {\bf q}}({\bf p};t)
  \hat {n}_{\bf q}({\bf p})-
  \int {\rd{\bf a}\,F({\bf a};t)\,\hat {f}({\bf a})}
\right\},
\end{equation}
where $ \rd{\bf a}\,\, \to \,\,\{\rd n_{\bf k}, \ \rd{\bf P}_{\bf k} , \
\rd\varepsilon_{\bf k}\}$.
The Massieu-Plank functional $\Phi(t)$ is determined from the normalization condition
$\mbox{Sp}\,\hat{\varrho}_\textrm{q}(t)\,=\,1:
$
\[
\Phi (t)= \ln \,\mbox{Sp}\,
\left\{ \exp \left[ - \sum\limits_{{\bf q,\,p}} \gamma_{-{\bf q}}({\bf p};t)
\hat {n}_{\bf q}({\bf p}) - \int \rd{\bf a}\,F({\bf a};t)\,\hat {f}({\bf a}) \right]\right\}.
\]
The functions $\gamma _{-{\bf q}}({\bf p};t)$ and $F({\bf a},t)$ are Lagrange multipliers and
can be defined from self-consistent conditions
\begin{equation}    \label{eq2.9}
f_1({\bf q},{\bf p};t)= \langle \hat{n}_{\bf q}({\bf p})\rangle^t =
\langle \hat{n}_{\bf q}({\bf p})\rangle_\textrm{q}^t\,, \qquad
f({\bf a};t)= \langle \hat{f}({\bf a})\rangle^t = \langle \hat{f}({\bf a})\rangle_\textrm{q}^t \,.
\end{equation}

The generalized solution of equation (\ref{eq2.3}) in  nonequilibrium statistical operator method
by Zubarev can be presented in the following form:

\begin{equation}             \label{eq2.10}
\hat{\varrho}(t) = \hat {\varrho }_\textrm{q} (t) -
\int\limits_{-\infty}^t \rd t'\,\re^{\varepsilon(t' - t)} \hat{T}_\textrm{q}
(t;t')\left[{1-P_\textrm{q}(t')}\right] \ri L_N \hat{\varrho }_\textrm{q} (t'),
\end{equation}
where
\begin{equation}               \label{eq2.11}
\hat{T}_\textrm{q}(t;t') = \exp_{+}
\left\{
-\int_{t'}^t \rd t' \left[{1-P_\textrm{q} (t')} \right]
\ri L_N
\right\}
\end{equation}
is the generalized time evolution operator that takes the projection into account.
To obtain the solution (\ref{eq2.10}) we used the Kawasaki-Gunton projection  operator
\cite{ZubMorRop1996.1,KawGunt1973} which in our case acts on the arbitrary statistical operator
according to the rule:
\begin{eqnarray}             \label{eq2.12}
\nonumber
 P_\textrm{q}(t)\hat{\varrho}'&=&\hat{\varrho}_\textrm{q}(t)\mbox{Sp}\hat{\varrho}' +
 \sum\limits_{{\bf q,\,p}}
 \frac{\partial \hat{\varrho}_\textrm{q}(t)}{\partial \langle \hat{n}_{\bf q} ({\bf p}) \rangle^t}
 \left\{\mbox{Sp}\left[ \hat{n}_{\bf q}({\bf p})\hat{\varrho}' \right] -
 \langle\hat{n}_{\bf q}({\bf p})\rangle^t \mbox{Sp}\hat{\varrho}'\right\}
      \\
 &&+\int{d{\bf a}}\frac{\partial\hat{\varrho }_\textrm{q}(t)}{\partial f({\bf a};t)}
 \left\{{\mbox{Sp}\left[{\hat{f}({\bf a})\hat{\varrho}'}\right]
 - f({\bf a};t)\mbox{Sp}\hat{\varrho}'}\right\}.
 \end{eqnarray}

We consider the action of the Liouville operator on the quasi-equilibrium operator (\ref{eq2.8}):
\begin{eqnarray}               \label{eq2.13}
 \ri L_N \hat{\varrho }_\textrm{q} (t)&=&
 -\sum\limits_{{\bf q,\,p}} {\gamma_{-{\bf q}} }({\bf p};t)
 \int\limits_{0}^{1} \rd\tau [\hat{\varrho}_\textrm{q}(t)]^{\tau}
 \dot{\hat{n}}_{\bf q}({\bf p})[\hat{\varrho}_\textrm{q}(t)]^{1-\tau} \nonumber
\\
 && - \int \rd{\bf a} F({\bf a};t) \int\limits_{0}^{1} \rd\tau[\hat{\varrho}_\textrm{q}(t)]^{\tau}
 \ri L_N \hat{f}(a)[\hat{\varrho}_\textrm{q}(t)]^{1-\tau},
\end{eqnarray}
where
$\dot{\hat{n}}_{\bf q}({\bf p}) = \ri L_N \hat{n}_{\bf q}({\bf p})$.
One introduces the operator function as in \cite{Moroz1981}
\begin{equation}                \label{eq2.14}
\hat J({\bf a})=\int \rd{\bf x} {\re}^{\ri{\bf x}(\hat{{\bf a}}-{\bf a})}
\int_0^1 \rd\tau {\re}^{-\ri\tau {\bf x}\hat{{\bf a}}} \ri L_N \hat{{\bf a}}
 \,{\re}^{\ri\tau {\bf x}\hat{\bf a}},
\end{equation}
then,
\begin{equation}                \label{eq2.15}
\ri\hat L_N \hat f({\bf a})=-\frac{\partial}{\partial{\bf a}}\hat J({\bf a})=
- \sum\limits^{5}_{m=1} \sum\limits_{\bf k}
  \frac{\partial \hat{J}_{m{\bf k}}({\bf a})}{\partial{\bf a}_{m{\bf k}}}\,.
\end{equation}
The second term in the right-hand side of (\ref{eq2.13}) can be represented as follows:
\begin{eqnarray}
\int \rd{\bf a} F({\bf a};t) \int\limits_{0}^{1}\rd\tau [\hat{\varrho}_\textrm{q}(t)]^{\tau}
\left(-\frac{\partial}{\partial{\bf a}}\hat{J}({\bf a})\right) [\hat{\varrho}_\textrm{q}(t)]^{1-\tau}=
\int \rd{\bf a} \left(\frac{\partial}{\partial{\bf a}}F({\bf a};t)\right) \int\limits_{0}^{1}\rd\tau
[\hat{\varrho}_\textrm{q}(t)]^{\tau} \hat{J}({\bf a})\left[\hat{\varrho}_\textrm{q}(t)\right]^{1-\tau}.
\nonumber
\end{eqnarray}
Now the expression (\ref{eq2.13}) reads
\begin{equation}             \label{eq2.16}
\ri L_N \hat\varrho_\textrm{q}(t)=-\sum\limits_{{\bf q,\,p}}
{\gamma_{-{\bf q}}}({\bf p};t)
\int\limits^{1}_{0} \rd\tau\, \dot{\hat{n}}_{\bf q}({\bf p};\tau) \hat{\varrho}_\textrm{q}(t)+
\int \rd{\bf a} \left(\frac{\partial}{\partial{\bf a}}F({\bf a};t)\right)
\int\limits^{1}_{0} \rd\tau\, \hat{J}({\bf a};\tau)\hat{\varrho}_\textrm{q}(t) ,
\end{equation}
where
\begin{equation}             \label{eq2.17}
\dot{\hat{n}}_{\bf q}({\bf p};\tau)=
[\hat{\varrho}_\textrm{q}(t)]^{\tau}\dot{\hat{n}}_{\bf q}({\bf p})
[\hat{\varrho}_\textrm{q}(t)]^{-\tau}\,,  \qquad
\hat{J}({\bf a};\tau)=
[\hat{\varrho}_\textrm{q}(t)]^{\tau}\hat{J}({\bf a})
[\hat{\varrho}_\textrm{q}(t)]^{-\tau}\,.
\end{equation}
Taking into account (\ref{eq2.16}) we represent the nonequilibrium statistical operator
(\ref{eq2.10}) in the following form:
\begin{eqnarray}         \label{eq2.18}
\hat{\varrho}(t) &=& \hat{\varrho}_\textrm{q}(t)+
\sum\limits_{\bf q,\,p} \int\limits^{t}_{-\infty}\rd t'
 {\re}^{\varepsilon (t' - t)}T_\textrm{q} (t,t')
 \left[ 1 - P_\textrm{q}(t') \right]
 \int\limits^{1}_{0} \rd\tau\, \dot{\hat{n}}_{\bf q}({\bf p};\tau)
 \gamma_{-\bf q}({\bf p};t')  \hat\varrho_\textrm{q}(t')          \nonumber       \\
&&+\int \rd{\bf a} \int\limits^{t}_{-\infty}\rd t'  {\re}^{\varepsilon (t' - t)}T_\textrm{q} (t,t')
 \left[ {1 - P_\textrm{q}(t')} \right]
 \int\limits^{1}_{0} \rd\tau\,\hat J({\bf a};\tau)
 \frac{\partial}{\partial{\bf a}}F({\bf a};t') \varrho_\textrm{q}(t')    \nonumber \\
&=& \hat\varrho_\textrm{q}(t)   +
 \sum\limits_{\bf q,\,p} \int\limits^{t}_{-\infty}\rd t'
 {\re}^{\varepsilon (t' - t)}T_\textrm{q} (t,t')
 \left[ {1 - P_\textrm{q}(t')} \right]
 \int\limits^{1}_{0} \rd\tau\,\dot{\hat{n}}_{\bf q}({\bf p};\tau)
 \gamma_{-\bf q}({\bf p};t')
 \hat\varrho_\textrm{q}(t')                \nonumber  \\
&& + \sum\limits_{\bf q} \int \rd {\bf a} \int\limits^{t}_{-\infty}\rd t'
 {\re}^{\varepsilon (t' - t)}T_\textrm{q} (t,t')
 \left[ {1 - P_\textrm{q}(t')} \right]     \nonumber  \\
&& \times \int\limits^{1}_{0} \rd\tau
\left\{
  \hat{J}_{n_{\bf q}}({\bf a};\tau)\frac{\partial}{\partial n_{\bf q}}F({\bf a};t')  +
  \hat{\bf J}_{{\bf P}_{\bf q}}({\bf a};\tau)\cdot\frac{\partial}{\partial {\bf P}_{\bf q}}F({\bf a};t') +  \hat{J}_{\varepsilon_{\bf q}}({\bf a};\tau)\frac{\partial}{\partial \varepsilon_{\bf q}}F({\bf a};t')
\right\}
    \varrho_\textrm{q}(t'),
\end{eqnarray}
which contains both nondissipative $\hat\varrho_\textrm{q}(t)$ and dissipative parts that
consistently describe non-Markovian kinetic and hydrodynamic processes with microscopic flows
$\dot{\hat{n}}_{\bf q}({\bf p};\tau)$, $\hat{J}({\bf a};\tau)$. Moreover,
\begin{eqnarray}             \label{eq2.19}
\dot{\hat{n}}_{\bf q}({\bf p})&=&-\ri[\hat{n}_{\bf q}({\bf p}),\hat{H}]_{-}=
 -\ri\frac{({\bf p}\cdot{\bf q})}{m}\hat{n}_{\bf q}({\bf p})   \nonumber\\
 && - \ri\frac{\sqrt{N}}{V} \sum\limits_{\bf k,\,p'}\nu(k)
\left(\delta_{\bf p',p-\frac{k}{2}}-\delta_{\bf p',p+\frac{k}{2}}\right)
  \hat{a}^{+}_{\bf p'+\frac{k-q}{2}} \hat{n}_{\bf k}
  \hat{a}_{\bf p'-\frac{k-q}{2}}
\end{eqnarray}
and one obtains the microscopic conservation law of particle density $\hat{n}_{\bf q}$:
\begin{equation}             \label{eq2.20}
\dot{\hat{n}}_{\bf q}=-\ri[\hat{n}_{\bf q},\hat{H}]_{-}=
-\ri({\bf q}\cdot \hat{\bf J}_{\bf q})=
-\frac{\ri}{m}({\bf q}\cdot \hat{\bf P}_{\bf q}),
\end{equation}
where $\hat{\bf J}_{\bf q}$ is the flow density operator of Bose particles.
Respectively, the expressions $J({\hat{\bf P}_{\bf q}};\tau)$,
$J({\hat{\varepsilon}_{\bf q}};\tau)$ contain the microscopic
conservation law of the momentum density operator
\begin{equation}             \label{eq2.21}
\dot{\hat{P}}^{\alpha}_{\bf q}=-\frac{\ri}{\sqrt{N}}\sum\limits_{\bf p} \frac{{p}_{\alpha}}{m}
    ({\bf p\cdot q})\hat{n}_{\bf q}({\bf p}) -
    \frac{\ri\sqrt{N}}{2V} \sum\limits_{\bf k}
    [\nu({\bf k})k_{\alpha} + \nu({\bf k+q})(-k_{\alpha}+q_{\alpha})]
\hat{n}_{\bf k}\hat{n}_{\bf -k+q}
\end{equation}
and the microscopic conservation law of the complete energy density operator
\begin{eqnarray}             \label{eq2.22}
\dot{\hat{\varepsilon}}_{\bf q}&=&-\frac{\ri}{\sqrt{N}}\sum\limits_{\bf p}
  \left(\frac{p^2}{2m}-\frac{q^2}{8m}\right) \frac{({\bf p \cdot q})}{m} \hat{n}_{\bf q}({\bf p})     \nonumber      \\
&&-\frac{\ri\sqrt{N}}{2V} \sum\limits_{{\bf k},\,\alpha}      \left\{
  \frac{\nu(\bf k)-\nu({\bf -k + q})}{2} k_{\alpha}  +
  \frac{\nu(\bf k)+\nu({\bf -k + q})}{2} q_{\alpha}   \right\}    \nonumber \\
&&   \times \frac{1}{m}\left[\hat{n}_{\bf k}, \hat{P}^{\alpha}_{\bf -k+q}\right]_{+}  +
\frac{1}{2V}\sum\limits_{\bf k} \nu(k)  \dot{\hat{n}}_{\bf q}.
\end{eqnarray}

The nonequilibrium statistical operator (\ref{eq2.18}) is a functional of the reduced description
parameters $f_1({\bf q,p};t)$ and $f(a;t)$ contained in self-consistent conditions
(\ref{eq2.9}) of determination of Lagrange multiplier $\gamma_{\bf q}({\bf p};t)$. The values
$F({\bf a};t)$ and $f_1({\bf q,p};t)$ are necessary for a complete description of transport
processes in a system. With this aim in mind, we use the conditions:
\[
\frac{\partial}{\partial t}\langle \hat{n}_{\bf q}({\bf p})\rangle^{t}=
\frac{\partial}{\partial t} f_{1}({\bf q,p};t) =
\langle \dot{\hat{n}}_{\bf q}({\bf p})\rangle^{t}, \qquad
\frac{\partial}{\partial t} f({\bf a};t)=\mbox{Sp}
\left( \hat{\varrho}(t)iL_{N}\hat{f}({\bf a}) \right).
\]
Calculating the average values in the right-hand parts of these equations with nonequilibrium
statistical operator (\ref{eq2.18}) we obtain a system of the following transport equations:
\begin{eqnarray}             \label{eq2.23}
\frac{\partial}{\partial t} f_1({\bf q,p};t) &+&
 \ri\frac{({\bf q \cdot p})}{m}  f_1({\bf q,p};t) =      
\frac{\ri\sqrt{N}}{V}\sum\limits_{{\bf k,\,p'}}  \nu(k)
  \left(\delta_{\bf p',p+\frac{k}{2}}-\delta_{\bf p',p-\frac{k}{2}}\right)
  f_{2}({\bf q,p,p',k};t) \nonumber  \\
&+& \sum\limits_{{\bf q',\,p'}}  \int\limits_{-\infty}^t \rd t'
  {\re}^{\varepsilon (t'-t)}
  \varphi_{nn}({\bf q, q',p,p'};t,t')\gamma_{{-\bf q'}}({\bf p}';t')    \nonumber \\
&-&  \int \rd{\bf a} \int\limits_{-\infty}^t \rd t'
 {\re}^{\varepsilon(t'-t)}
 \varphi_{nJ}({\bf q,p,};t,t')\frac{\partial}{\partial{\bf a}} F({\bf a};t'),
 \end{eqnarray}
\begin{eqnarray}             \label{eq2.24}
\nonumber \frac{\partial}{\partial t} f({\bf a};t) &+&
\frac{\partial}{\partial{\bf a}} \langle \hat{J}({\bf a}) \rangle^{t}_\textrm{q} =    
-\sum\limits_{{\bf q',\, p'}}  \int\limits_{-\infty}^t \rd t'
  {\re}^{\varepsilon (t'-t)}  \frac{\partial}{\partial{\bf a}}
  \varphi_{Jn}({\bf a},{\bf q',p'};t,t')\gamma_{{-\bf q'}}({\bf p}';t')       \\
&+& \int \rd{\bf a}' \int\limits_{-\infty}^t \rd t'
 {\re}^{\varepsilon(t'-t)}  \frac{\partial}{\partial{\bf a}}
 \varphi_{JJ}({\bf a,a'};t,t')\frac{\partial}{\partial {\bf a}'} F({\bf a}';t'),   \qquad\qquad
 \end{eqnarray}
where
\begin{equation}             \label{eq2.25}
 f_{2}({\bf q,p,p',k};t) = \mbox{Sp}
\left\{
 \hat{a}^{+}_{\bf p'+\frac{k-q}{2}} \hat{n}_{\bf q}
 \hat{a}_{\bf p'-\frac{k-q}{2}}  \hat{\varrho}_\textrm{q} (t)
\right\}
\end{equation}
is the two-particle quasi-equilibrium distribution function of Bose particles,
 \begin{equation}             \label{eq2.26}
\varphi_{nn}({\bf q,q',p,p'};t,t') = \langle
 \hat{I}_{n}({\bf q,p};t)\, \hat{T}_\textrm{q}(t,t')
 \int\limits^{1}_{0}\rd\tau \,\hat{I}_{n}({\bf q',p'};t',\tau)
\rangle^{t'}_\textrm{q}.
\end{equation}
 is the transport kernel (memory function)
which describes the dissipation of kinetic processes.
Quantities $\varphi_{nJ}({\bf q,p},{\bf a}';t,t')$, $\varphi_{Jn}({\bf a},{\bf q,p};t,t')$ are
the matrices with elements
\begin{eqnarray}             \label{eq2.27}
 \varphi_{nJ_{l}}({\bf q,p},a';t,t') =
  \langle  \hat{I}_{n}({\bf q,p};t)\, \hat{T}_\textrm{q}(t,t')
  \int\limits^{1}_{0}\rd\tau \,\hat{I}_{J_{l}}({\bf a}';t',\tau)
\rangle^{t'}_\textrm{q}\,,                    \nonumber  \\
 \varphi_{J_{l}n}({\bf a},{\bf q,p};t,t') =
 \langle  \hat{I}_{J_{l}}({\bf a};t) \, \hat{T}_\textrm{q}(t,t')
 \int\limits^{1}_{0}\rd\tau \,\hat{I}_{n}({\bf q,p};t',\tau)
\rangle^{t'}_\textrm{q}\,.
\end{eqnarray}
These elements are the transport kernels which describe the dissipation between kinetic and
hydrodynamic processes. $\varphi_{JJ}(a,a';t,t')$ is the matrix with elements
\begin{equation}             \label{eq2.28}
 \varphi_{J_{l} J_{f}}({\bf a,a}';t,t') = \langle
   \hat{I}_{J_{l}}({\bf a};t) \, \hat{T}_\textrm{q}(t,t')
   \int\limits^{1}_{0}\rd\tau \,\hat{I}_{J_{f}}({\bf a}';t',\tau)
 \rangle^{t'}_\textrm{q},
\end{equation}
which describe the dissipation of hydrodynamic processes in a quantum Bose fluid. Transport kernels
(\ref{eq2.26})--(\ref{eq2.28}) are constructed on generalized flows
\begin{equation}             \label{eq2.29}
  \hat{I}_{n}({\bf q,p};t) = Q(t) \dot{\hat{n}}_{\bf q}({\bf p}), \qquad
  \hat{I}_{J_{l}}({\bf a};t) = Q(t) \hat{J}_{l}({\bf a}),
\end{equation}
where $Q(t)=1-P(t)$. Operator $P(t)$ is the generalized projection Mori operator
which acts on any operator $\hat{A}$ according to the rule
\[
\nonumber
 P(t)\hat{A}=\langle\hat{A}\rangle^{t}_\textrm{q} + \sum\limits_{\bf q,\, p}
 \frac{\partial\langle\hat{A}\rangle^{t}_\textrm{q}}{\partial \langle\hat{n}_{\bf q}({\bf p})\rangle^{t}}
  \left(\hat{n}_{\bf q}({\bf p})-\langle\hat{n}_{\bf q}({\bf p})\rangle^{t}   \right)  +
  \int d{\bf a}\frac{\partial\langle\hat{A}\rangle^{t}_\textrm{q}}{\partial \langle\hat{f}({\bf a}) \rangle^{t}}\left(\hat{f}({\bf a})-\langle\hat{f}({\bf a})\rangle^{t}  \right)
\]
and corresponds to the structure of the projection Kawasaki-Gunton operator $P_\textrm{q}(t)$ (\ref{eq2.12}).
It is important to note that the transport kernels contain both a contribution from quantum diffusion
in the coordinate and momentum space and a contribution from the generalized function ``force-force''.
One can readily derive this by substituting (\ref{eq2.19}) into $\varphi_{nn}({\bf q,q',p,p'};t,t')$
and open the contribution from kinetic and potential parts of (\ref{eq2.19}):
\begin{eqnarray}             \label{eq2.30}
\nonumber
 \varphi_{nn}({\bf q,p,q',p'};t,t')&=&\frac{1}{m^{2}}
 {\bf q}\cdot D_{nn}({\bf q,p,q',p'};t,t')\cdot {\bf q'}-
  \frac{1}{m}{\bf q}\cdot{D}_{nF}({\bf q,p,q',p'};t,t')      \\
  &&-{D}_{Fn}({\bf q,p,q',p'};t,t')\frac{1}{m}{\bf q'} +
  {D}_{FF}({\bf q,p,q',p'};t,t'),
\end{eqnarray}
where
\begin{equation}             \label{eq2.31}
{D}_{nn}({\bf q,p,q',p'};t,t')=\langle Q(t) \cdot{\bf p}\hat{n}_{\bf q}({\bf p})
  \cdot \hat{T}_\textrm{q}(t,t') Q(t') \cdot{\bf p'}\hat{n}_{\bf q}({\bf p'}) \rangle_\textrm{q}^{t}
\end{equation}
is the generalized diffusion coefficient of quantum particles in the space
$\{{\bf q,p}\}$.
Other components have the following structure:
\begin{eqnarray}             \label{eq2.32}
\nonumber
{D}_{nF}({\bf q,p,q',p'};t,t')&=&\langle Q(t)\cdot {\bf p}\hat{n}_{\bf q}({\bf p})\cdot \hat{T}_\textrm{q}(t,t') \,Q(t') \cdot F_{\bf q'}({\bf p'}) \rangle_\textrm{q}^{t}\,,              \\
{D}_{Fn}({\bf q,p,q',p'};t,t')&=&\langle Q(t) \cdot F_{\bf q}({\bf p})\cdot \hat{T}_\textrm{q}(t,t')
    \,Q(t') \cdot {\bf p'}\hat{n}_{\bf q'}({\bf p'}) \rangle_\textrm{q}^{t}\,,   \nonumber \\
{D}_{FF}({\bf q,p,q',p'};t,t')&=&\langle Q(t)\cdot F_{\bf q}({\bf p})\cdot \hat{T}_\textrm{q}(t,t')
    \,Q(t') \cdot F_{\bf q'}({\bf p'})     \rangle_\textrm{q}^{t}\,,
\end{eqnarray}
where
$$
F_{\bf q}({\bf p})=
\frac{\ri\sqrt{N}}{V} \sum\limits_{\bf k,\,p'} \nu(k)
  \left(\delta_{\bf p',p+\frac{k}{2}} -\delta_{\bf p',p-\frac{k}{2}}\right)
  \hat{a}^{+}_{\bf p'+\frac{k-q}{2}} \hat{n}_{\bf q}({\bf p})
 \hat{a}_{\bf p'-\frac{k-q}{2}}\,.
$$

For a detailed study of the mutual effect of kinetic and hydrodynamic processes we will allocate
the ``kinetic'' part in quasi-equilibrium statistical operator using the operator representation:
\begin{equation}             \label{eq2.33}
\hat{\varrho}_\textrm{q}(t) = \hat{\varrho}^\textrm{k}_\textrm{q}(t) -\int \rd a F({\bf a};t)
\int\limits^1_0 \rd\tau U(F|\tau)\hat{f}({\bf a};\tau)\hat{\varrho}^\textrm{k}_\textrm{q}(t),
\end{equation}
where
\begin{equation}             \label{eq2.34}
 \hat{\varrho}^\textrm{k}_\textrm{q}(t) = \mbox{exp}
 \left\{-\Phi^\textrm{k}(t) - \sum\limits_{\bf q,\,p}\gamma_{-\bf q}({\bf p};t)\hat{n}_{\bf q}({\bf p})\right\},  \qquad
 \Phi^\textrm{k}(t) = \ln \,\mbox{Sp}\, \exp
 \left\{-\sum\limits_{{\bf q,\,p}}\gamma_{-{\bf q}}({\bf p};t)\hat{n}_{\bf q}({\bf p}) \right\}
\end{equation}
is the quasi-equilibrium statistical operator which is the basis of the kinetic level of description,
and
\begin{equation}         \label{eq2.35}
\hat{f}({\bf a};\tau)=[\hat{\varrho}^\textrm{k}_\textrm{q}(t)]^{\tau}
      \hat{f}({\bf a})[\hat{\varrho}^\textrm{k}_\textrm{q}(t)]^{-\tau}.
\end{equation}
The operator $U(F|\tau)$ satisfies the equation
\begin{equation}             \label{eq2.36}
 U(F|\tau) = 1 -\int d{\bf a} F({\bf a};t) \int\limits^{\tau}_0 \rd \tau' U(F|\tau')
                   \hat{f}({\bf a};\tau').
\end{equation}
We use the expression (\ref{eq2.33}) for  determining the Lagrange multiplier $F(a;t)$ from the
self-consistent condition:
\begin{equation}             \label{eq2.37}
 f({\bf a};t)=\langle \hat{f}({\bf a}) \rangle^{t}_\textrm{q}=\langle\hat{f}({\bf a})\rangle^{t}_\textrm{k}-
 \int \rd{\bf a}' W({\bf a,a}';t,\tau)F({\bf a}';t),
\end{equation}
where
\begin{equation}             \label{eq2.38}
 W({\bf a,a}';t) = \int\limits^1_0 \rd\tau  \langle
  \hat{f}({\bf a}) U(F|\tau)  \hat{f}({\bf a}';\tau)
 \rangle^{t}_\textrm{k}
\end{equation}
is the structure function, in which the averaging is implemented with quasi-equilibrium statistical
operator (\ref{eq2.34}). From (\ref{eq2.37}) we find $F(a;t)$:
\begin{equation}             \label{eq2.39}
F({\bf a};t) = -\int \rd{\bf a}' \delta f({\bf a}';t)W_{-1}({\bf a,a}';t),
\end{equation}
where
\begin{equation}             \label{eq2.40}
\delta f({\bf a};t) = f({\bf a};t) - \langle \hat{f}({\bf a}) \rangle^{t}_\textrm{k} =
       \langle \hat{f}({\bf a})\rangle^{t} - \langle \hat{f}({\bf a}) \rangle^{t}_\textrm{k}
\end{equation}
are the fluctuations of the distribution function of hydrodynamic variables determined as a difference
between the complete distribution function and the one averaged with operator
$\hat{\varrho}_\textrm{q}^\textrm{k}(t)$. Taking into account equation (\ref{eq2.39}) the quasi-equilibrium
statistical operator $\hat{\varrho}_\textrm{q}(t)$ can be written as follows:
\begin{equation}             \label{eq2.41}
\hat{\varrho}_\textrm{q}(t) = \hat{\varrho}^\textrm{k}_\textrm{q}(t) + \int \rd{\bf a} \int \rd{\bf a}'
 \int\limits^{1}_{0} \rd\tau U(F|\tau) W_{-1}({\bf a',a};t)
 \hat{f}({\bf a};\tau) \delta f({\bf a}';t)
 \hat{\varrho}^\textrm{k}_\textrm{q}(t).
\end{equation}
The function $W_{-1}({\bf a,a}';t)$ is the inverse to the structure function $W({\bf a,a}';t)$ and is the
solution of the integral equation:
\begin{equation}             \label{eq2.42}
\int \rd{\bf a}'' W({\bf a,a}'';t)W_{-1}({\bf a}'',{\bf a}';t) = \delta ({\bf a-a}').
\end{equation}
Only the functions $W({\bf a,a}';t)$ and $W_{-1}({\bf a,a}';t)$ satisfy the equation (\ref{eq2.42})
which have singular parts:
\begin{equation}             \nonumber
 W({\bf a,a}';t) = W({\bf a};t)[\delta({\bf a-a}') + R({\bf a,a}';t)], \quad
 W_{-1}({\bf a,a}';t) = W_{-1}({\bf a};t)[\delta({\bf a-a}')+r({\bf a,a}';t)],
\end{equation}
where $R({\bf a,a}';t)$ and $r({\bf a,a}';t)$ are the regular parts and
\[
W({\bf a};t)=\int \rd{\bf a}'\,W({\bf a,a}';t), \qquad
  W_{-1}({\bf a};t)=\int \rd{\bf a}'\,W_{-1}({\bf a,a}';t).
\]
An important point is that the expression (\ref{eq2.39}) is the equation to determine $F({\bf a};t)$,
because the function $W_{-1}({\bf a,a}';t)$ depends on $F({\bf a};t)$ according to the structure
function $W({\bf a,a}';t)$, which in turn depends on $U(F|\tau)$ (\ref{eq2.36}).

By means of (\ref{eq2.39}) and (\ref{eq2.41}) we rewrite the equation system (\ref{eq2.23}),
(\ref{eq2.24}) in the following form:
\begin{eqnarray}             \label{eq2.43}
\nonumber
\frac{\partial}{\partial t} f_1({\bf q,p};t) &+&
\ri\frac{({\bf q \cdot p})}{m}  f_1({\bf q,p};t) =
\frac{\ri\sqrt{N}}{V}\sum\limits_{{\bf k,\,p'}}  \nu(k)
  \left(\delta_{\bf p',p+\frac{k}{2}}-\delta_{\bf p',p-\frac{k}{2}}\right)
  f_{2}({\bf p,q,p',k};t)    \\
&+ & \sum\limits_{{\bf q',\,p'}}  \int\limits_{-\infty}^t \rd t'
  {\re}^{\varepsilon (t'-t)}
  \varphi_{nn}({\bf q, q',p,p'};t,t')\gamma_{{-\bf q'}}({\bf p}';t') \nonumber\\
&+&\int \rd{\bf a}' \int \rd{\bf a}'' \int\limits_{-\infty}^t \rd t'
 {\re}^{\varepsilon(t'-t)}
 \varphi_{nJ}({\bf q,p,};t,t')\frac{\partial}{\partial {\bf a}'}
   W_{-1}({\bf a}',{\bf a}'';t') \delta f({\bf a};t'),
\end{eqnarray}
\begin{eqnarray}             \label{eq2.44}
\nonumber
\frac{\partial}{\partial t} \delta f({\bf a};t)&-&
\sum\limits_{{\bf q',p'}}\Omega_{fn}({\bf a,q',p'};t)\gamma_{-\bf q'}({\bf p}';t') +
\frac{\partial}{\partial{\bf a}} \int \rd{\bf a}' \int \rd{\bf a}'' \upsilon({\bf a,a}'';t)
    W_{-1}({\bf a}'',{\bf a}';t)\delta f({\bf a}';t)     \\ \nonumber
=&-&\sum\limits_{{\bf q',\, p'}}  \int\limits_{-\infty}^t \rd t'
  {\re}^{\varepsilon (t'-t)}  \frac{\partial}{\partial{\bf a}}
  \varphi_{Jn}({\bf a},{\bf q',p'};t,t')\gamma_{{-\bf q'}}({\bf p}';t')  \nonumber  \\
&-&\int \rd{\bf a}' \int \rd{\bf a}'' \int\limits_{-\infty}^t \rd t'
 {\re}^{\varepsilon(t'-t)}  \frac{\partial}{\partial{\bf a}}
 \varphi_{JJ}({\bf a,a}';t,t')\frac{\partial}{\partial{\bf a}'} W_{-1}({\bf a}',{\bf a}'';t')
 \delta f({\bf a}'';t'),
 \end{eqnarray}
where the generalized hydrodynamic velocities
\begin{equation}             \label{eq2.45}
 \upsilon({\bf a,a}';t) = \int \rd{\bf a}'' \int\limits^1_0 \rd\tau \mbox{Sp}\left\{ \hat{J}({\bf a})
  Q(F|\tau) \hat{f}({\bf a}'';\tau) \hat{\varrho}^\textrm{k}_\textrm{q}(t)   \right\}
  W_{-1}({\bf a}'',{\bf a}';t).
\end{equation}
were introduced.
While deriving these equations, the following property was used:
$$\frac{\partial}{\partial{\bf a}}\langle \hat{J}({\bf a}) \rangle^{t}_\textrm{k}=
 -\frac{\partial}{\partial t}\langle \hat{f}({\bf a}) \rangle^{t}_\textrm{k}-
 \sum\limits_{{\bf q',\, p'}}\Omega_{fn}(a;{\bf q',\, p'};t)\gamma_{{-\bf q'}}({\bf p}';t'),$$
where
\[
\Omega_{fn}(a;{\bf q',p'};t)=\mbox{Sp}\left\{\hat{f}({\bf a})\int\limits_{0}^{1}\rd\tau
[\hat{\varrho}^\textrm{k}_\textrm{q}(t)]^{\tau} I_{n}^\textrm{k}({\bf q',p'};t) [\hat{\varrho}^\textrm{k}_\textrm{q}(t)]^{1-\tau}\right\}
\]
is the time correlation function between $\hat{f}({\bf a})$ and $I_{n}^\textrm{k}({\bf q',p'};t)$.
Here, the generalized kinetic flows \linebreak $I_{n}^\textrm{k}({\bf
q,p};t)=[1-P^\textrm{k}(t)]\dot{\hat{n}}_{\bf q}({\bf p})$ and the kinetic generalized projection Mori
operator
\[P^\textrm{k}(t)\hat{A}=\langle\hat{A}\rangle_\textrm{k}^{t}+\sum\limits_{{\bf q,p}}\frac{\delta
\langle\hat{A}\rangle_\textrm{k}^{t}}{\delta \langle \hat{n}_{\bf q}({\bf p})\rangle^{t}}\left(\hat{n}_{\bf
q}({\bf p})-\langle \hat{n}_{\bf q}({\bf p})\rangle^{t}\right)
\]
were introduced.

The two limiting cases follow from the system of transport equations (\ref{eq2.43}),
(\ref{eq2.44}). First, unless we consider the nonlinear hydrodynamic correlations we will obtain
the kinetic equation for Wigner function of quantum Bose particles. Second, if we do not take into account
the kinetic processes, then we will obtain a Fokker-Plank equation for distribution function
$f({\bf{a}};t)$, that corresponds to the results of the article \cite{Moroz1981}:
\begin{eqnarray}             \label{eq2.46}
 \frac{\partial}{\partial t}f({\bf a};t) &+&  \frac{\partial}{\partial{\bf a}}
 \int \rd{\bf a}' \upsilon({\bf a,a}') f({\bf a}';t) =       \nonumber       \\
&-&
\int\limits^t_{-\infty} \rd t' {\re}^{\varepsilon(t'-t)}
  \frac{\partial}{\partial{\bf a}} \int \rd{\bf a}' K({\bf a,a}';t-t')
  \frac{\partial}{\partial{\bf a}'} \int \rd{\bf a}'' W_{-1}({\bf a',a''})
  f({\bf a}'';t),
\end{eqnarray}
where (\ref{eq2.45}) transforms to $\upsilon ({\bf a,a}')$ from \cite{Moroz1981}
\begin{equation}             \label{eq2.47}
 \upsilon({\bf a,a}') = \int \rd{\bf a}'' \mbox{Sp}\left\{
 \hat{J}({\bf a})\hat{f}({\bf a}'')W_{-1}({\bf a}',{\bf a}'')\right\}
\end{equation}
and the structure function is reduced to
\begin{equation}             \label{eq2.48}
 W({\bf a,a}') = \mbox{Sp} \left\{ \hat{f}({\bf a})\hat{f}({\bf a}') \right\}.
\end{equation}
 Accordingly,  $K({\bf a,a}';t)$ is the matrix with elements
\begin{equation}            \label{eq2.49}
 K_{lf}({\bf a,a}';t) = \mbox{Sp}\left\{\hat{I}_{l}({\bf a})\hat{T}_\textrm{q}(t,t')\hat{I}_{f}({\bf a}')
                 \right\},
\end{equation}
where $\hat{I}_{l}({\bf a})=(1-P)\hat{J}_{l}({\bf a})$ are the dissipative flows with
projection operator
\[
P\hat{A}=\int \rd{\bf a} \int \rd{\bf a}' \hat{f}({\bf a})W_{-1}({\bf a,a}')
\mbox{Sp}\left\{ \hat{A}\hat{f}({\bf a}') \right\}.
\]

If we neglect the memory effects on the hydrodynamical level
in the system of transport equation (\ref{eq2.43}), (\ref{eq2.44}),
then we will obtain the following system of equations:
\begin{eqnarray}             \label{eq2.50}
\frac{\partial}{\partial t} f_1({\bf q,p};t)
&+& \ri\frac{({\bf q \cdot p})}{m}  f_1({\bf q,p};t) \nonumber \\
&=&
\frac{\ri\sqrt{N}}{V}  \sum\limits_{{\bf k,\,p'}}  \nu(k)
  \left(\delta_{\bf p',p+\frac{k}{2}}-\delta_{\bf p',p-\frac{k}{2}}\right)
  f_{2}({\bf p,q,p',k};t)  \nonumber \\
\nonumber
&+& \sum\limits_{{\bf q',\,p'}}  \int\limits_{-\infty}^t \rd t'
  {\re}^{\varepsilon (t'-t)}
  \varphi_{nn}({\bf q, q',p,p'};t,t')\gamma_{{-\bf q'}}({\bf p}';t')\nonumber \\
 & -&
\int \rd{\bf a}' \varphi_{nJ}({\bf q,p,};{\bf a}')\frac{\partial}{\partial{\bf a}'}
   W_{-1}({\bf a}';t) \delta f({\bf a};t),
 \end{eqnarray}
\begin{eqnarray}             \label{eq2.51}
\frac{\partial}{\partial t} \delta f({\bf a};t) &+&
\frac{\partial}{\partial a}v({\bf a};t) \delta f({\bf a};t)  \nonumber \\
=& -&\sum\limits_{{\bf q'\, p'}}   \frac{\partial}{\partial{\bf a}}
  \varphi_{Jn}({\bf a},{\bf q',p'})\gamma_{{-\bf q'}}({\bf p}';t) -
\frac{\partial}{\partial{\bf a}}
 \varphi_{JJ}({\bf a})\frac{\partial}{\partial{\bf a}}W_{-1}({\bf a};t)\delta f({\bf a};t),
 \end{eqnarray}
in which the memory effects on the kinetic level are preserved.
In this system, the following designations are used:
\begin{equation}             \label{eq2.52}
 \varphi_{n J_{l}}({\bf q,p,a}) = \int\limits_{-\infty}^{0} \rd t
 {\re}^{\varepsilon t}
 \langle  \hat{I}_{n}({\bf q,p}) \hat{T}_\textrm{q}(t) \hat{I}_{J_{l}}({\bf a})
 \rangle^{t}_\textrm{k}\,,
\end{equation}
\begin{equation}             \label{eq2.53}
 \varphi_{J_{l}J_{f}}({\bf a}) = \int \rd{\bf a}' \int\limits_{-\infty}^{0} \rd t
 {\re}^{\varepsilon t}
 \langle  \hat{I}_{J_{l}}({\bf a}) \hat{T}_\textrm{q}(t) \hat{I}_{J_{f}}({\bf a}')
 \rangle^{t}_\textrm{k}\,,
\end{equation}
where $v({\bf a};t)$ is the contribution from a singular part of generalized velocity $v({\bf a},{\bf
a}';t)= v({\bf a};t)\delta({\bf a-a}')+u({\bf a,a}',t)$, during which $v({\bf a};t)=\langle J({\bf
a};\tau)\rangle_\textrm{k}^{t}$.
Another point to emphasize is that a contribution only from a singular part of the structure
function $W_{-1}({\bf a,a}';t)$ is present in the equation system, namely $W_{-1}({\bf a};t)$. Such local
approximation can be used near the critical point when the values that strongly fluctuate are
hydrodynamical variables and long-wave components of the order parameter.

A hard problem to examine the nonlinear fluctuations based on the equation system
(\ref{eq2.43}), (\ref{eq2.44}) is to calculate the structure function $W(a,a';t)$ and
generalized hydrodynamical velocities $v({\bf a};t)$. To this end, we use the method of iterations. In
the first approximation for the operator function $U(F|\tau)$,  we take
$$U^{(1)}({F|\tau})=1,$$
which follows from (\ref{eq2.36}), and then by substiting into (\ref{eq2.38}), we obtain the first
approximation for the structure function:
\begin{equation}             \label{eq2.54}
 W^{(1)}({\bf a,a}';t) = \int\limits^1_0 \rd\tau  \langle
  \hat{f}({\bf a}) \hat{f}({\bf a}';\tau) \rangle^{t}_\textrm{k}\,.
\end{equation}
In a similar  manner using the following approximation
\[
U^{(2)}({F|\tau})=-\int \,\rd{\bf a} F({\bf a};t)
  \int\limits^{\tau}_0 \rd\tau'\,\hat{f}({\bf a};\tau')
\]
for the second approximation of the structure function, one obtains
\begin{equation}             \label{eq2.55}
 W^{(2)}({\bf a,a}';t) = -\int\,\rd{\bf a}'' F({\bf a}'',t)
 \int\limits^1_0 \,\rd\tau \int\limits^{\tau}_0 \rd\tau'
 \langle \hat{f}({\bf a}) \hat{f}({\bf a}'';\tau) \hat{f}({\bf a}';\tau) \rangle^{t}_\textrm{k}\,.
\end{equation}
In a such a manner, the structure function is equal to
\begin{equation} \label{eq2.56}
  W({\bf a,a}';t)= W^{(1)}({\bf a,a}';t)+W^{(2)}({\bf a,a}';t)+\ldots \, .
\end{equation}

To calculate the first approximation of the structure function, we present it as follows:
\begin{equation}             \label{eq2.57}
 W^{(1)}({\bf a,a}';t) = W^{(1)}({\bf a};t)\left[\delta({\bf a-a}') + R^{(1)}({\bf a,a}';t)\right],
\end{equation}
where
\begin{equation}             \nonumber
W^{(1)}({\bf a};t) = \int \rd{\bf a'}\,W^{(1)}({\bf a,a'};t) =
   \langle \hat{f}(a) \rangle^{t}_\textrm{k}\,.
\end{equation}
After simple transformations, for the regular part of the structure function we can obtain \cite{Moroz1981}
%
\begin{equation}             \nonumber
R^{(1)}({\bf a,a'};t) = \int \rd{\bf x}\int \rd{\bf x'}
  {\re}^{\ri{\bf x'a' }-\ri{\bf xa }} \int^{1}_{0}d\tau
  \langle {\re}^{\ri{\bf x}\hat{{\bf a}}} {\re}^{-\ri{\bf x'}\hat{{\bf a}}'(\tau)} -
  {\re}^{\ri({\bf x-x'})\hat{{\bf a}}} \rangle^{t}_\textrm{k}
  \frac{1}{W^{(1)}({\bf a};t)}\,.
\end{equation}
From this expression it follows that if the basis operators commute with one another, then $R({\bf a,a'};t)$
vanishes. Thus, the presence of the regular part of  $W({\bf a,a'};t)$ (reciprocally in function
$W_{-1}({\bf a,a'};t)$) is characteristic only of a quantum system.

Now we calculate the function $W^{(1)}({\bf a};t)$
\begin{equation}             \label{eq2.58}
W^{(1)}({\bf a};t)=\int \rd{\bf x} {\re}^{-\ri{\bf xa}}\,\mbox{Sp}\left\{
   {\re}^{\ri{\bf x}\hat{{\bf a}}}\,\hat{\varrho}^\textrm{k}_\textrm{q}(t)
\right\}.
\end{equation}
Using the cumulant expansion, we write it in the following form:
\begin{equation}              \label{eq2.59}
 W^{(1)}({\bf a};t)=\int \rd{\bf x}
 \exp\left\{-\ri\sum\limits_{\alpha}x_{\alpha}a_{\alpha}+
 \sum\limits_{l=1}\frac{(\ri)^l}{l!}\sum\limits_{\alpha_{1} \ldots \alpha_{l}}
  x_{\alpha_{1}} \ldots x_{\alpha_{l}}
 M_{\alpha_{1} \ldots \alpha_{l}} \right\},
\end{equation}
where $\alpha = \{m, \bf{k}\}$ and $\sum_{\alpha}(\ldots)=\sum^{5}_{m=1}\sum_{\bf{k}}(\dots)$.
To calculate cumulants $M_{\alpha_1 \ldots \alpha_l}$, we write the average from the exponent as follows:
\begin{eqnarray*}
\langle {\re}^{\sum_{\alpha} \ri x_{\alpha}\hat{a}_{\alpha}}\rangle^t_\textrm{k} &=&
 \sum\limits^{\infty}_{l=0}\frac{{\ri}^l}{l!}\sum\limits_{\alpha_{1}\ldots\alpha_{l}}
    x_{\alpha_{1}} \ldots x_{\alpha_{l}}
    \langle a_{\alpha_{1}}\ldots a_{\alpha_{l}} \rangle^{t}_\textrm{k}\\
    &=&
 \exp\left\{\sum^{\infty}_{l=1}\frac{\ri^l}{l!}\sum_{\alpha_{1}\ldots\alpha_{l}}
    x_{\alpha_{1}} \ldots x_{\alpha_{l}}M_{\alpha_{1}\ldots\alpha_{l}}\right\}
\end{eqnarray*}
and expand the right-hand side in a series. We compare the coefficients at the same products
of $x_{\alpha}$ and for the first three cumulants obtain:
\begin{eqnarray}                 \label{eq2.60}
\nonumber
 M_{\alpha}&=&\langle \hat{a}_{\alpha} \rangle^{t}_\textrm{cum}=
    \langle \hat{a}_{\alpha} \rangle^{t}_\textrm{k}\,,    \\
 M_{\alpha_{1}\alpha_{2}}&=& \langle
 \hat{a}_{\alpha_{1}}\hat{a}_{\alpha_{2}}\rangle^{t}_\textrm{cum} =
 \langle
 \hat{a}_{\alpha_{1}}\hat{a}_{\alpha_{2}}\rangle^{t}_\textrm{k}-
 \langle \hat{a}_{\alpha_{1}}\rangle^{t}_\textrm{k}\langle \hat{a}_{\alpha_{2}}
 \rangle^{t}_\textrm{k}\,,                 \nonumber        \\
\nonumber
 M_{\alpha_{1}\alpha_{2}\alpha_{3}} &=& \langle
 \hat{a}_{\alpha_{1}}\hat{a}_{\alpha_{2}}\hat{a}_{\alpha_{3}}\rangle^{t}_\textrm{cum} =
 \langle \hat{a}_{\alpha_{1}}\hat{a}_{\alpha_{2}}\hat{a}_{\alpha_{3}}\rangle^{t}_\textrm{k}-
 \frac{3}{2}\langle \hat{a}_{\alpha_{1}}\rangle^{t}_\textrm{k}
   \langle \hat{a}_{\alpha_{2}} \hat{a}_{\alpha_{3}}  \rangle^{t}_\textrm{k} \nonumber \\
&&-
 \frac{3}{2}\langle \hat{a}_{\alpha_{1}}\hat{a}_{\alpha_{2}}\rangle^{t}_\textrm{k}
   \langle \hat{a}_{\alpha_{3}} \rangle^{t}_\textrm{k} +
 2\langle \hat{a}_{\alpha_{1}}\rangle^{t}_\textrm{k}
  \langle \hat{a}_{\alpha_{2}}\rangle^{t}_\textrm{k}
  \langle \hat{a}_{\alpha_{3}}\rangle^{t}_\textrm{k}\,,
\end{eqnarray}
which are averaged with quasi-equilibrium statistical operator $\hat{\varrho}_\textrm{q}^\textrm{k}(t)$.

Now, we separate the sum over $l$ in the exponent (\ref{eq2.59}) in two parts: with $l\leqslant2$ and
$l\geqslant3$. Thus, we select a Gaussian component and  expand the rest in a series. As a result, we
have
\begin{equation}                \label{eq2.61}
 W^{(1)}({\bf a};t)=\int \rd{\bf x} \,
 \exp\left\{\ri\sum\limits_{\alpha}x_{\alpha}(M_{\alpha} - a_{\alpha})
    -\frac{1}{2}\sum\limits_{\alpha_1\alpha_2} x_{\alpha_1}x_{\alpha_2}M_{\alpha_1\alpha_2}\right\}
 \left\{1 + \Lambda + \frac{1}{2}\Lambda^2 + \frac{1}{3!}\Lambda^3+ \ldots
 \right\},
\end{equation}
where
\begin{equation}            \nonumber
 \Lambda =
 \sum\limits_{l\geqslant 3} \frac{\ri^{l}}{l!}\sum\limits_{\alpha_{1}\alpha_{2}} \ldots
  \sum\limits_{\alpha_{l}} x_{\alpha_{1}}x_{\alpha_{2}}\ldots x_{\alpha_{l}}
  M_{\alpha_{1}\alpha_{2}\ldots \alpha_{l}}\,.
\end{equation}
Let us consider the Gaussian approximation
\begin{equation}                \label{eq2.62}
 W_{G}({\bf a};t)=\int \rd{\bf x}
\exp\left\{\ri\sum\limits_{\alpha}x_{\alpha}(M_{\alpha} - a_{\alpha})
    -\frac{1}{2}\sum\limits_{\alpha_1\alpha_2} x_{\alpha_1}x_{\alpha_2}M_{\alpha_1\alpha_2}\right\}
\end{equation}
and integrate it with respect to $\bf x$. To this end, we should transform the quadratic form
$x_{\alpha_1}x_{\alpha_2}M_{\alpha_1\alpha_2}$ in the exponent into a diagonal form. We need to
solve the equation for the eigenvalues and eigenvectors of the symmetric matrix $M_{\alpha_1\alpha_2}$
\begin{equation}                \nonumber
 \sum\limits_{\alpha_2} M_{\alpha_1\alpha_2}v_{\alpha\alpha_2}=z_{\alpha}v_{\alpha\,\alpha_1},
  \qquad
 \det \mid M_{\alpha_1\alpha_2}- z_{\alpha}{\bf I} \mid =0,
\end{equation}
where  $v_{\alpha\,\alpha_1}$, $z_{\alpha}$ are the eigenvectors and eigenvalues,
$\bf I$ is the unit matrix. Then, out of the eigenvectors we construct the transition matrix
$v_{\alpha_{1} \alpha_{2}}$ to the new variables ${\bf y}$ where the coefficient matrix is diagonal
and consists of eigenvalues $\bar{Q}_{\alpha}=\bar{M}_{\alpha \alpha_{1}}=\delta_{\alpha \alpha_{1}}
z_{\alpha}$. Old variables $\bf x$ are connected with the new ones by the relation
$x_{\alpha}=\sum\limits_{\alpha_{1}}\,v_{\alpha \alpha_{1}}y_{\alpha_{1}}$. Now, the exponent in
(\ref{eq2.62}) has the form of perfect squares
$\sum\limits_{\alpha_{1}\alpha_{2}}x_{\alpha_1}x_{\alpha_2}M_{\alpha_1\alpha_2} =
\sum\limits_{\alpha} \bar{Q}_{\alpha}y^2_{\alpha}$
and can be integrated with respect to the variables $\bf y$.

\section{Concluding remarks}

The nonequilibrium statistical operator was obtained for a consistent description of kinetic and nonlinear
hydrodynamic fluctuations in a quantum Bose system. In order to consider the kinetic
processes, the nonequilibrium one-particle Wigner function is used as a parameter in a reduced
description. The distribution function of hydrodynamic variables is chosen for the study of nonlinear
hydrodynamic fluctuations.
While deriving a quasi-equilibrium statistical operator, the ``kinetic'' part
$\hat{\varrho}^\textrm{k}_\textrm{q}(t)$ and the part connected with the superoperator $U(F|\tau)$,
that leads to the approximation in terms of correlation functions for $\hat{f}(a)$,
were examined in detail.
The equation of the Fokker-Plank type was obtained, which is related to the kinetic equation.
The nonequilibrium distribution function makes it possible to calculate the averaged values of
$\langle\hat{a}_{l} \ldots \hat{a}_{l}\rangle^t= \int \rd a \, \left(a_{l} \ldots a_{l}\right)f(a;t)$,
and to obtain operators $\hat{\bf J}_{\bf q}$ for flux density and Reynolds-type
chain of equations for values $\langle \hat{\bf J}_{\bf q}\rangle^t$, $\langle \hat{\bf J}_{\bf q}\,\hat{\bf
J}_{\bf q'}\rangle^t$, $\langle \hat{\bf J}_{\bf q}\hat{\bf J}_{\bf q'}\hat{\bf J}_{\bf
q''}\rangle^t$ as in classical case \cite {Zubar1982}.
This result is important for the study of quantum turbulence phenomena \cite{QunTurb2009}.

We have considered the first two approximations for $U^{(1)}(F|\tau)$ and
$U^{(2)}(F|\tau)$, which allowed us to obtain the structure of function $W(a,a';t)$ given by
Eq.~(\ref{eq2.56}). We proposed a method to calculate the structure function in the first approximation
for $W^{(1)}(a,a';t)$ using the cumulant expansion (approximation with respect to correlations) with a
Gaussian distribution for collective variables. Similar calculations can be used for
hydrodynamic velocities (\ref{eq2.45}) which is important for microscopic derivation of
the generalized transport kernels (\ref{eq2.28}). Such a calculation of the structure function enables
us to consider the Fokker-Plank equation in Gauss approximations and higher, 
and to obtain a chain of Reynolds-type equations for time correlation functions
$\langle \hat{a}_{l}\ldots\hat{a}_{j}\rangle^{t}$.
The generalized hydrodynamical velocities $v({\bf a};t)$
will be calculated in the same manner as in the classical case
\cite{Zubar1982,IdzIgnTok1996,MorTokIdzKob1996} using the cumulant representations in Gauss
approximations or higher. Then, the transport kernels in the Fokker-Plank equation in
the mode-coupling-like form could be presented similarly to the classical case \cite{Ignat1999}.
The case of quantum statistics requires a special consideration which will be presented in a subsequent paper.

\ukrainianpart

\title{Квантові рівняння переносу для бозе-систем з врахуванням нелінійних
       гідродинамічних процесів}

\author{П.А.~Глушак, М.В.~Токарчук}
\address{Інститут фізики конденсованих систем НАН України, вул. І.~Свєнціцького, 1,
 79011 Львів,  Україна}

\makeukrtitle

\begin{abstract}

Використовуючи метод нерівноважного статистичного оператора Зубарєва, запропоновано підхід для опису кінетики з врахуванням нелінійних гідродинамічних флуктуацій для квантової бозе-системи.
Розраховано нерівноважний статистичний оператор, що узгоджено описує як кінетичні, так і нелінійні гідродинамічні процеси. Отримано кінетичне рівняння для нерівноважної одночастинкової функції
розподілу та узагальнене рівняння Фоккера-Планка для гідродинамічних змінних (густин імпульсу, енергії і кількості частинок). В кумулянтному наближенні розраховано структурну функцію гідродинамічних флуктуацій. Це надає можливість проаналізувати узагальнене рівняння Фоккера-Планка в гаусовому і вищих наближеннях для динамічних кореляцій, що важливо для опису квантових турбулентних процесів.
\keywords бозе-система, гелій, кінетика, гідродинаміка, кореляційна функція, рівняння
Фоккера-Планка
\end{abstract}

\begin{thebibliography}{46}

\bibitem{ZubarMoroz1984} Zubarev D.N., Morozov V.G.,  Theor. Math. Phys., 1984,
    \textbf{60}, No.~2, 814; \doi{10.1007/BF01018982}.

\bibitem{ZubMorOmelTok1991} Zubarev~D.N., Morozov~V.G., Omelyan~I.P., Tokarchuk~M.V.,
    Theor. Math. Phys., 1991, \textbf{87}, No.~1, 412; \\ \doi{10.1007/BF01016582}.

\bibitem{ZubMorOmelTok1993} Zubarev~D.N., Morozov~V.G., Omelyan~I.P., Tokarchuk~M.V.,
    Theor. Math. Phys., 1993, \textbf{96}, No.~3, 997; \\ \doi{10.1007/BF01019063}.

\bibitem{TokOmelKobr1998} Tokarchuk M.V., Omelyan I.P., Kobryn O.E., Condens. Matter
    Phys., 1998, \textbf{1}, 687; \doi{10.5488/CMP.1.4.687}.

\bibitem{Zubar1974} Zubarev~D.N., Non-equilibrium Statistical Thermodynamics,
    New York, Consultant Bureau, 1974.

\bibitem{ZubarMoroz1989} Zubarev D.N., Morozov V.G.,
In: Collection of scientific works of Mathematical Institute of USSR Academy of Sciences,
Vol.~{191}, Nauka, Moscow, 1989, p. 140 (in Russian).

\bibitem{ZubMorRop1996.1} Zubarev~D.N., Morozov~V.G., R\"opke~G., Statistical Mechanics
    of Non-equilibrium Processes, Vol.~1. Basic Concepts, Kinetic Theory,
    Akademie Verlag-Wiley VCN, Berlin, 1996.

\bibitem{ZubMorRop1996.2} Zubarev~D.N., Morozov~V.G., R\"opke~G., Statistical
    Mechanics of Non-equilibrium Processes, Vol.~2. Relaxation and Hydrodynamics Processes,
    Akademie Verlag-Wiley VCN, Berlin,  1996.

\bibitem{PethcSmith2008} Pethick C.J., Smith H., Bose-Einstein Condensation in Dilute Gases, Cambridge University Press, Cambridge, 2008.

\bibitem{GrifNikZar2009} Griffin A., Nikuni T., Zaremba E., Bose Condensed Gases at Finite
    Temperatures, Cambridge University Press, Cambridge, 2009.

\bibitem{Khalat1961} Khalatnikov I.M., Theory of Superfluidity,
   Nauka, Moscow, 1971 (in Russian).

\bibitem{Patter1974} Patterman S., Superfluid Hydrodynamics, North-Holland, Amsterdam,
 1974.

\bibitem{Glyde1994} Glyde H.R., Exitations in Liquid and Solid Helium, Clarendon
    Press, Oxford, 1994.

\bibitem{Griff1993} Griffin A., Exitations in a Bose-condensed Liquid. Cambridge University Press, Cambridge,     1993.

\bibitem{KovalPelet2006} Kovalevsky M.Yu., Peletminskii S.V., Statistical Mechanics of Quantum Liquids and Crystals,  Fizmatlit, Moscow, 2006 (in Russian).


\bibitem{Moroz1983} Morozov V.G., Physica A, 1983, \textbf{117}, 511; \doi{10.1016/0378-4371(83)90129-2}.

\bibitem{Moroz1986} Morozov V.G., Theor. Math. Phys., 1986, \textbf{67}, No.~1, 404; \doi{10.1007/BF01028894}.

\bibitem{LebSukhKhal1981} Lebedev V.V., Sukhorukov A.I., Khalatnikov I.M.,
    J. Exp. Theor. Phys., 1981, \textbf{53}, No.~4, 733.

\bibitem{KovalPeletSlus1982} Kovalevsky~M.Yu., Lavrinenko~N.M., Peletminsky~S.V., Sokolovsky~A.I.,
Theor. Math. Phys., 1982, \textbf{50}, No.~3, 296; \doi{10.1007/BF01016462}. 

\bibitem{Moroz1981} Morozov V.G., Theor. Math. Phys., 1981, \textbf{48}, No.~3, 807; \doi{10.1007/BF01019317}.

\bibitem{AkhezPelet1981} Akhiezer A.I., Peletminsky S.V., Methods of Statistical
    Physics,  Pergamon Press, New York, 1981.

\bibitem{KirkpDorf1985.1} Kirkpatrick T.R., Dorfman I.R., J. Low Temp. Phys.,
    1985, \textbf{58}, No.~3/4, 301; \doi{10.1007/BF00681309}.

\bibitem{KirkpDorf1985.2} Kirkpatrick T.R., Dorfman I.R., J. Low Temp. Phys.,
    1985, \textbf{58}, No.~5/6, 399; \doi{10.1007/BF00681133}.

\bibitem{KirkpDorf1985.3} Kirkpatrick T.R., Dorfman I.R., J. Low Temp. Phys.,
    1985, \textbf{59}, No.~1/2, 1; \doi{10.1007/BF00681501}.

\bibitem{ZarGrifNik1998} Zaremba E., Griffin A., Nikuni T., Phys. Rev. A, 1998, \textbf{57}, 4695; \doi{10.1103/PhysRevA.57.4695}.

\bibitem{ZarNikGrif1999} Zaremba E.,  Nikuni T., Griffin A., J. Low Temp. Phys., 1999,
    \textbf{116}, No.~3/4, 277; \doi{10.1023/A:1021846002995}.



\bibitem{LauckVasconLuzzi1990} Lauck~L., Vasconcellos~\'A.R., Luzzi~R., Physica A, 1990,
    \textbf{168}, 789; \doi{10.1016/0378-4371(90)90031-M}.

\bibitem{MadurVasconLuzzi1998.1} Madureira J.R., Vasconcellos~\'A.R., Luzzi~R., J. Chem. Phys.,     1998, \textbf{108}, 7568; \doi{10.1063/1.476191}.

\bibitem{MadurVasconLuzzi1998.2} Madureira J.R., Vasconcellos~\'A.R., Luzzi~R., J. Chem. Phys.,
    1998, \textbf{108}, 7580; \doi{10.1063/1.476192}.

\bibitem{Tserkov1990} Tserkovnikov Yu.A., Theor. Math. Phys., 1990, \textbf{85}, 1096; \doi{10.1007/BF01017252}.
%
\bibitem{Tserkov1990_2} Tserkovnikov Yu.A., Theor. Math. Phys., 1990, \textbf{85}, 1192; \doi{10.1007/BF01086849}.     

\bibitem{Tserkov1992.1995} Tserkovnikov Yu.A., Theor. Math. Phys., 1992, \textbf{93}, 1367; \doi{10.1007/BF01016395}. 
%
\bibitem{Tserkov1992.1995_2} Tserkovnikov Yu.A., Theor. Math. Phys. 1995, \textbf{105}, 1249; \doi{10.1007/BF02067493}.  

\bibitem{Zubarev1960} Zubarev D.N., Sov. Phys. Usp., 1960, \textbf{3}, 320; \doi{10.1070/PU1960v003n03ABEH003275}.

\bibitem{Tserkov1986} Tserkovnikov Yu.A., Theor. Math. Phys., 1986, \textbf{69}, 1254; \doi{10.1007/BF01017624}. 

\bibitem{VakHlTok1997} Vakarchuk I.O., Hlushak P.A., Tokarchuk M.V., Ukr. Fiz. Zh., 1997, \textbf{42}, 1150 (in Ukrainian).

\bibitem{HlTok2005} Hlushak P.A., Tokarchuk M.V., Condens. Matter Phys., 2004,
    \textbf{7}, No.~3(39), 639; \doi{10.5488/CMP.7.3.639}.

\bibitem{Onuki2004} Onuki A., Phase Transition Dynamics, Cambridge University Press, Cambridge, 2004.

\bibitem{QunTurb2009} Progress in Low Temperature Physics: Quantum Turbulence, vol.~XVI, Tsubota~M.,  Halperin W.P. (Eds.),  Elsevier, Amsterdam, 2006.

\bibitem{Nemirov2006} Nemirovskii S.K., Phys. Rep., 2013, \textbf{524}, 85; \doi{10.1016/j.physrep.2012.10.005}.

\bibitem{ResibLen1977} R\'esibois~P., de~Leener~M., Classical Kinetic Theory of
    Fluids, John Willey \& Sons, New York, 1977.

\bibitem{Kawasaki1976} Kawasaki K., In: Phase Transition and Critical Phenomena, Vol.~5A,
Domb C., Green M.S. (Eds.), Academic, New York,  1976, p. 165--411.

\bibitem{ZubarMoroz1983} Zubarev D.N., Morozov V.G., Physica A, 1983, \textbf{120},
    No.~3, 411; \doi{10.1016/0378-4371(83)90062-6}.

\bibitem{PeletSlus1992} Peletmiskii S.V., Slusarenko Yu.V., Prob. At.
    Sci. Technol., 1992, \textbf{3}(24), 145 (in Russian).

\bibitem{KawGunt1973} Kawasaki K., Gunton J.D., Phys. Rev. A, 1973, \textbf{8}, 2048; \doi{10.1103/PhysRevA.8.2048}.

\bibitem{Zubar1982} Zubarev D.N., Theor. Math. Phys., 1982, \textbf{59}, No.~1, 1004; \doi{10.1007/BF01014797}.

\bibitem{IdzIgnTok1996} Idzyk I.M., Ighatyuk V.V., Tokarchuk M.V., Ukr. Fiz. Zh.,
    1996, \textbf{41}, No.~10, 1017 (in Ukrainian).

\bibitem{MorTokIdzKob1996} Morozov V.G., Tokarchuk M.V., Idzyk I.M., Kobryn A.E.,
    Preprint of the Institute for Condensed Matter Physics, ICMP–96–15U, Lviv, 1996 (in Ukrainian).

\bibitem{Ignat1999} Ignatyuk V.V., Condens. Matter Phys., 1999, \textbf{2}, No.~1(17), 37; \doi{10.5488/CMP.2.1.37}.


\end{thebibliography}
\end{document}